\documentclass{aa}
\usepackage{graphics}
\begin{document}

      \thesaurus{06     
              (08.03.3
               08.01.2
               08.18.1
               08.05.3
                10.15.2)} 
   \title{Ca II activity and rotation in F-K evolved
   stars\fnmsep\thanks{Based on observations collected at ESO, La
   Silla. Tables 1-3 are only available in electronic form
at the CDS via anonymous ftp to cdsarc.u-strasbg.fr (130.79.128.5)
or via http://cdsweb.u-strasbg.fr/Abstract.html}}

   \subtitle{ }

   \author{L. Pasquini\inst{1} \and
          J.R. de Medeiros\inst{2} \and L. Girardi\inst{3,4} }

   \offprints{L. Pasquini}

   \institute{ European Southern Observatory,
               Karl-Schwarzschild-Strasse 2, 85748 Garching bei M\"unchen \\
              email: lpasquin@eso.org
         \and
             Departamento de F\'isica, Universidade Federal do Rio Grande do 
Norte, 59072-970, Natal, RN, Brazil\\
             email: renan@dfte.ufrn.br
          \and 
              Max-Planck-Institut f\"ur Astrophysik, Karl-Schwarzschild-Strasse 1, 85748 Garching bei M\"unchen              
          \and
           Present address: Dipartimento di Astronomia, Vicolo dell' 
Osservatorio 5, I-35122 Padova, Italy
}

   \date{Received : Accepted}

   \maketitle

   \begin{abstract}

Ca II H and K high resolution observations for 60 evolved stars in the 
field and in 5 open clusters  are presented. 
From these spectra chromospheric fluxes 
are derived, and a {\it homogeneous} sample of more than 100 giants 
is built adding data from the literature. 
In addition, for most stars, rotational velocities
were derived from CORAVEL observations. 
By comparing chromospheric emission in the cluster stars
we confirm the 
results of Pasquini and Brocato (1992):  chromospheric activity 
depends on the stellar effective temperature,
{\it and} mass, when intermediate mass stars ($M\sim4 M_\odot$) are 
considered. 
The Hyades and the Praesepe clump 
giants show  the same level of activity, as expected from stars 
with similar masses and effective temperatures.  
A difference of up to 0.4 dex in the chromospheric fluxes among the  
Hyades giants is recorded and this sets a clear limit to the 
intrinsic spread of stellar activity in evolved giants. 
These differences in otherwise very similar stars are likely due to 
stellar cycles and/or differences in  the stellar 
initial angular momentum.
Among the field stars none of the giants with (V-R)$_o < $0.4 
and Ia supergiants  observed shows  a signature of Ca II activity; 
this can be  due either to the  real absence of a chromosphere, 
but also to other causes which preclude the appearance of Ca II reversal. 

By analyzing the whole sample we find that 
chromospheric activity scales linearly with stellar rotational velocity and 
a high power of stellar effective temperature: 
F'$_k \propto T_{\rm eff}^{7.7}$ (Vsini)$^{0.9}$. 
This result can be interpreted as the effect of two chromospheric 
components of different nature: one mechanical and one magnetic. 

Alternatively, by using the Hipparcos parallaxes and evolutionary tracks, 
we divide the sample according to the stellar masses, and we follow 
the objects along an evolutionary track. 
For each range of masses activity can simply be expressed as a function of 
{\it only one parameter}: either the T$_{eff}$ or the  
  angular rotation $\Omega$, 
with laws F'$_k \sim$ $\Omega ^{\alpha}$, because angular velocity decreases 
with effective temperature along an evolutionary track.  

By using the evolutionary tracks and the observed {\it Vsini} we 
investigate the evolution of the angular momentum for evolved stars in the 
range 1-5 M$_{\odot}$. For the 1.6-3 solar mass stars the data are 
consistent with the I$\Omega$=const law while lower and higher masses 
follow a law similar to  
I$\Omega ^2$=const,
where I is the computed 
stellar momentum of inertia. We find it intriguing that {\it Vsini} 
remains almost constant for 1M$_{\odot}$ stars along  their evolution; 
if a similar behavior is shared by Pop II stars, this could explain the 
relatively high degree of activity observed in Pop II giants. 

Finally, through the use of models, we have verified the consistency 
of the F'$_k \propto \Omega ^{\alpha}$ and the I$\Omega ^{\beta}$ = Const
laws derived, finding an excellent agreement. 

This representation, albeit crude (the models do not consider, 
for instance, mass losses) 
represents the evolution of Ca II activity and of the
angular momentum  in a satisfactory way in most of the portion of 
H-R diagram analyzed. 
Different predictions could be tested  with observations 
in selected clusters.
 
   \end{abstract}

%

\section{Introduction}

The study of activity in the external layers of evolved stars  is a 
quite exciting topic, which still needs firm answers to a number of 
questions. We do not really know  how these chromospheres 
are powered, and what is the precise relationship
between the presence of chromospheres and mass losses, with the 
interesting implications this may have for stellar evolution 
(see e.g. Cacciari and Freeman 1983, Dupree et al. 1992). 

In main sequence stars stellar activity is linked to 
stellar rotational velocity and to the evolution of the angular momentum 
(i.e. Pallavicini et al. 1981, Noyes et al. 1984); this last quantity, 
often neglected in stellar evolutionary theory,
 is becoming more and more relevant in several aspects, 
for instance when extra mixing mechanisms are required 
to explain the evolution of elements 
like Li, Be and C (see, i.e. 
Zahn 1992, Charbonnel and Do Nascimento 1998). 
In general, by studying both chromospheres and angular momentum evolution 
of evolved stars we also aim at better understanding  convection, 
which is one of the 
most critical aspects in the modeling of the stellar interior.

In this work we have derived
chromospheric fluxes based on high resolution spectroscopy in the K line of Ca 
II. In addition 
for almost all stars we have derived stellar rotational velocities; 
we also use the Hipparcos 
parallaxes to place the stars in the H-R diagram 
and we finally use  theoretical evolutionary models 
to estimate stellar masses and to interpret the observed data. 
In addition to stars in the field we have 
observed 15 giants belonging to 5 open clusters of different turn-off masses, 
and we  use these data to guide the interpretation of  the field stars.


\section{Observations and Sample}

The  Ca II observations were obtained at ESO, La Silla
 in several observing runs in the years 1990-1994. 
 Most observations were obtained at the ESO  1.52m with the Echelec 
spectrograph (Lindgren 1989), while CASPEC at the ESO 3.6m 
(Randich and Pasquini 1997) was used to observe the 
giants in the open clusters IC 4651 and NGC 6067; for both 
configurations the resolving power obtained was similar, 
with R$\sim$22000. 

Data reduction was performed by using the echelle 
context of the MIDAS package; special care was 
taken in observing  several early type stars, to correct 
properly for the blaze function of the spectrographs, using the recepies  
described in Pasquini (1992). In the case of the CASPEC observations 
of NGC 6067, a quite reddened cluster,  one  early type 
star belonging to the cluster  was observed. 
This spectrum allowed us  to eliminate  
the strong IS Ca II K line which would  otherwise have heavily 
contaminated our spectra. 
The spectrum of the early star  was fitted, leaving  the narrow 
interstellar absorption line 
at its position. By dividing the 
late type star spectra by the early type one, the IS Ca II is
therefore eliminated. 
One example is shown in Figure 1, where the spectrum of the 
giant CPD-537416 
is shown after the correction for the IS line (continuous line) 
and before (dashed line). 
The strong 
IS line, if not considered, would not only alter the chromospheric flux of the 
cool giants, but also the K line core asymmetry.


   \begin{figure}
\resizebox{\hsize}{!}{\includegraphics{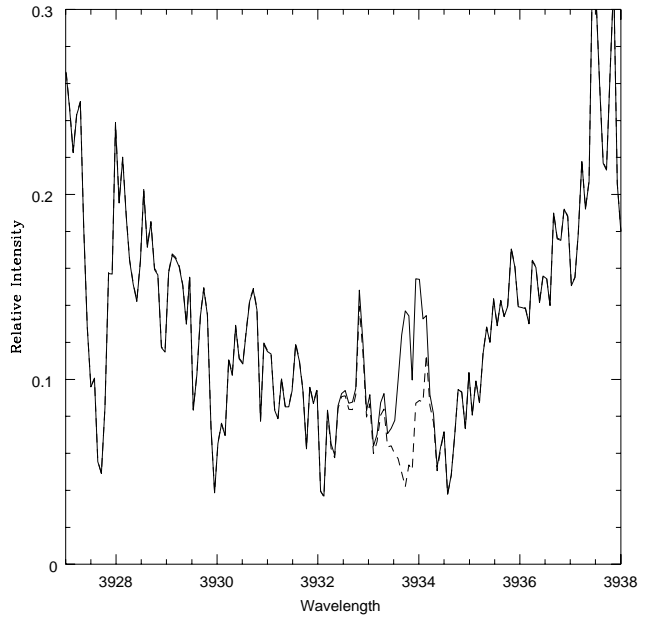}}
\caption{ The Ca II core spectrum of the giant
CPD-537416 in NGC6067 is shown, before the correction for the IS 
Ca II (dashed line) and after (continuous line). Note the change in  
the asymmetry of the line core}
         \label{Fig1}%
    \end{figure}

The full sample analyzed in this work 
consists of three different groups: 

\begin{itemize}
\item Stars published by Pasquini and Brocato (1992, here-after PB92);
for these stars CORAVEL rotational 
 velocities  and Hipparcos parallaxes are now available.
\item 15 Stars belonging to 5 clusters, selected with the aim of 
 covering the  largest 
possible range of ages and turn-off masses. In addition, we 
also make use of the M67 data from Dupree et al. 1999. 
\item 45 new Ca II observations of field giants, all of them with 
measured CORAVEL rotational velocities.
   
\end{itemize}

The sample is not complete in any aspect.  Most stars were 
selected because of their 
bright apparent magnitude, in addition in the new observations 
several hot giants were observed, because 
this region in the H-R diagram was not well covered in previous studies. 
We emphasize, however, that stars were selected neither on the basis
of their previously known  
rotational velocity nor  activity level; 
we should not have therefore inserted any bias with respect to these 
quantities. 

\section{Data Analysis}

For the sample stars, we aim at determining several quantities: 
chromospheric fluxes, rotational velocities, 
stellar effective temperatures and absolute magnitudes. 

In addition, for our analysis it is essential to have estimates of the stellar
masses. To this end we make use of several evolutionary tracks  of solar 
metallicity computed
with the updated Padua code (Girardi et al. 2000). The momentum of inertia 
 I has been computed for each of the evolutionary points along the tracks. 
For this study, no mass losses were considered; we note however that 
for the low and intermediate mass stars we do not expect mass losses to 
modify substantially I (and therefore our results) at least if 
a standard Reimers (1975) mass loss law is assumed (see section 4.4 
below).

For all stars Ca II K line chromospheric fluxes are computed following
the calibrations of 
Pasquini et al. (1988) and Linsky et al. (1979),  
and effective temperatures are derived for all stars following the 
calibration of Pasquini et al. (1990). More recent T$_{eff}$
temperature calibrations exist, nevertheless the 
Pasquini et al. (1990) calibration 
has the advantage that it is based on  the V-R (Johnson)
color, which is the same used to compute the  stellar chromospheric fluxes; 
since most of our analysis is based on relative scales, 
this will help in reducing  systematic effects.
Note also that the effective temperatures derived are used exclusively to 
infer stellar masses and rotational periods;  changes in the temperature scale 
may therefore have some impact on the correlation coefficient derived,
but they will not  influence our main conclusions. 

Absolute magnitudes were derived by using Hipparcos parallaxes, 
reddening is relevant only for 
some bright giants and supergiants (Pasquini et al. 1990); 
e(V-R) was estimated 
on the basis of Johnson (1966), then  the law 
e(B-V)=0.66 e(V-R) was applied and finally absolute 
visual magnitudes were derived by using the 
standard absorption law A(V)=3.1 e(B-V).

Since from our spectra Wilson-Bappu (1957) magnitudes can be derived 
and these magnitudes  were used in the previous, pre-Hipparcos studies, 
we have computed Wilson-Bappu M$_v$ for all stars, 
and compared them  with the Hipparcos ones. 
Most of the observed stars are within 200 pc, and 
their Hipparcos parallaxes are therefore very accurate. 
In Figure 2 the comparison 
between absolute magnitude as computed by using the Wilson-Bappu 
law following Lutz (1970) (as applied in PB92) 
and the Hipparcos parallaxes is shown. 
A systematic trend is present:
the W-B law gives systematically brighter 
magnitudes than Hipparcos. We think that a re-calibration 
of the Wilson Bappu law with Hipparcos stars would become a powerful
tool  to infer the 
absolute magnitude of late type stars.


   \begin{figure}
\resizebox{\hsize}{!}{\includegraphics{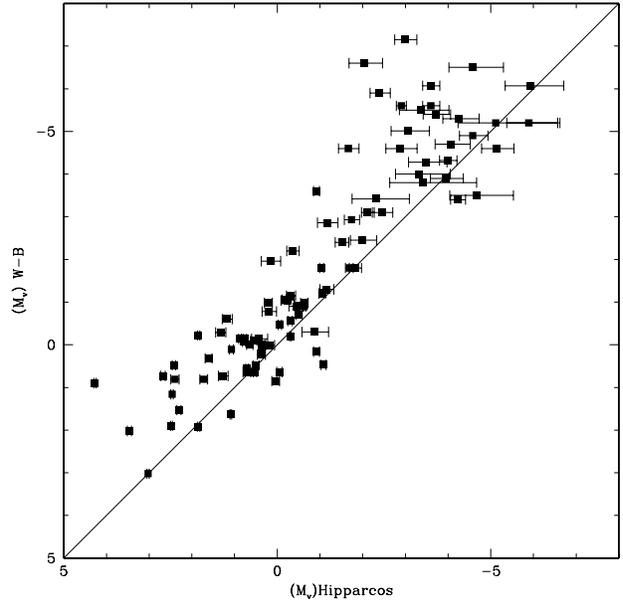}}
\caption{ Comparison between the Visual
         absolute magnitudes as derived from Hipparcos and from the 
   Wilson Bappu law, following Lutz (1970) }%
         \label{Fig2}%
    \end{figure}

As far as rotational velocities are concerned, the  {\it Vsini} measurements 
were derived from the CORAVEL survey of de Medeiros and Mayor (1999), 
but several stars have {\it Vsini} measured by Gray and coworkers (Gray 1982, 
1983, Gray and Pallavicini 1989, Gray and Toner 1987, Gray and Nagar 1985);
the claimed accuracy for both analyses is in the range of  0.5-1 km/sec.
Since for $\sim$ 50 stars both CORAVEL and Gray rotational velocities
are available, a comparison  can be performed, and it is shown in Figure 3.  
For most stars the agreement between the two sets of measurements is 
excellent, within 1 km/sec, but discrepancies much larger than the quoted 
errors exist: the results are consistent with the results of De
Medeiros  and Mayor (1999) who found for a larger sample of giants 
and subgiants: {\it Vsini}(COR)= -1.15 + 1.18$\times${\it Vsini}(Gray);
since  the  CORAVEL measurements are available for all stars, we will 
adopt them in the following, ensuring an homogeneous data set.

   \begin{figure}
\resizebox{\hsize}{!}{\includegraphics{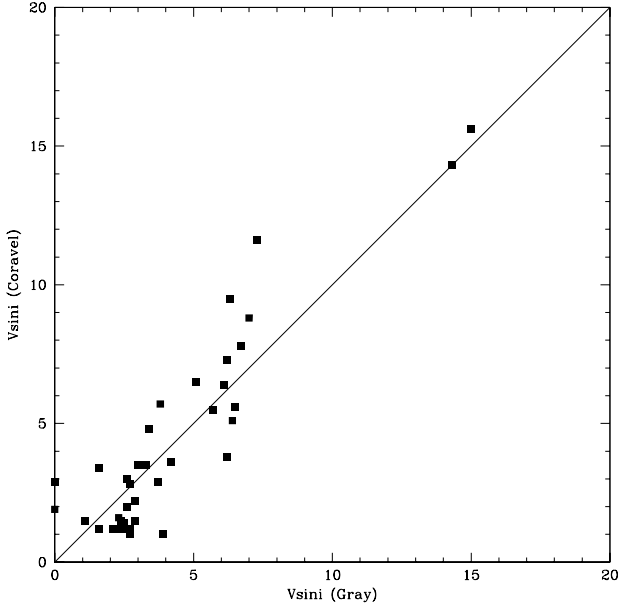}}
\caption{Comparison between CORAVEL and
         Gray {\it Vsini} for the sample stars common to the two 
         groups }
         \label{Fig3}%
    \end{figure}

\subsection{Previously Published stars}

PB92 investigated the dependence of Ca II activity 
from several stellar parameters, using Ca II 
observations of quality comparable 
with the present one, and 
using the same calibration, V-R temperature scale, 
bolometric corrections by Flower (1977), 
and determining the stellar reddening following Johnson(1966). 

Table 1 contains the updated PB92 data. The differences from PB92 are
that absolute magnitudes are now 
based on the Hipparcos parallaxes and the CORAVEL rotational velocities 
have been added, as well as a 
binarity flag, which indicates if the stars are likely single (S), 
spectroscopic binary (SB) and  with known 
orbit (SBO).  
 
From the PB92 sample, for only 5 stars 
{\it Vsini} measurements 
were unavailable (all of spectral class M) and these 
stars are not considered in the present subsample,  which therefore
consists of 60 stars. Rotational velocities from Gray and collaborators 
are given, when available.  
G refers to Gray (1982, 1983), 
GN to Gray and Nagar (1985), GT to Gray and Toner 
(1987), GP to Gray and Pallavicini (1989).

 \subsection{Cluster stars} 

In addition to the field stars, vital 
information may be obtained  by analyzing 
stars belonging to open clusters, for which basic stellar 
parameters are better constrained. 
As part of this project, 15 giants  belonging to 5 clusters were
observed; the clusters  are (ordered as
decreasing turn-off masses): NGC 2516, NGC 6067,  Praesepe, Hyades, 
and IC4651. 
The cluster stars were analyzed in the same way as the field stars, 
and the same effective temperature scale was adopted. 

In Table 2, the data on the observed clusters are presented; 
turn-off masses, reddening and distance modulus are taken from
Meynet and Mermilliod (1993) for 
NGC 2516 (E(B-V)=0.12, turn-off=4.3M$_{\odot}$), NGC 6067 
(E(B-V)=0.314, turn-off=4.5M$_{\odot}$) and IC 4651 (E(B-V)=0.08, turn-off=
1.4 M$_{\odot}$).
For Hyades and Praesepe, distances 
are derived directly from Hipparcos parallaxes; 
E(B-V)=0  and turn-off=2.05 M$_{\odot}$ have been assumed.
 A$_k$ refers to the ratio between the Ca II K line core area 
and the pseudocontinuum at 3950{\AA}
(Pasquini et al. 1988), while A(50) 
refers to the same area, divided by the 3925-3975 bandpass 
(Linsky et al. 1979).

\subsection{New Field stars}

The last subsample includes the new Echelec 
observations of field stars. These observations were acquired 
to enlarge the PB92 sample, both to obtain  more stars 
with measured rotational velocities and to 
explore Ca II activity among bright (and hot) giants and supergiants, 
a region of the H-R diagram poorly investigated  
in previous studies. 

Table 3 summarizes  the basic data for this subsample; column names are as in 
Table 1, while A$_k$ and A(50) are as in Table 2. 
It is worth noticing that for 12 stars no signature of Ca II K line core 
reversal is observed. All these stars are either rather hot 
(with (V-R)$_o \le 0.4$), or belonging to 
the Ia supergiant class. 
We will return to this point in the discussion.

\section{Discussion}

\subsection{Cluster Stars}

The obvious advantage of observing stars in clusters is to eliminate the 
uncertainties always present 
in the stellar parameters of field stars. 
Stars belonging to a cluster can for most purposes be  approximated 
to a homogeneous sample in age, chemical composition, and, 
when referring to the evolved stars,
masses as well.  In addition, when observing one cluster, the magnitude and 
temperature sequence of 
the stars is usually well defined, at least on a {\it relative} scale. 

We recall that, since the cluster star observations are 
processed in the same way as the 
field stars, they form an homogeneous sample with them; this implies that it 
should be possible
to extend the results of the cluster stars  to the field ones. 

In Figure 4  Ca II fluxes are shown as a function of the 
stellar effective temperature. Stars from different clusters are marked with 
different symbols. 
In this Figure, in addition to the data of Table 2, 
the sequence of giants from M67 (Dupree et al. 1999) is added. 
The M67 data were acquired at a similar spectral 
resolution and analyzed in a similar
way to the present sample. 
A different Ca II calibration was used, and when 
the difference between the calibrations is taken into account, the results are
in excellent agreement (Dupree et al. 1999); 
 the M67 Ca II fluxes  from Dupree et al. (1999) have been corrected 
to be homogeneous with the present analysis. 

   \begin{figure}
\resizebox{\hsize}{!}{\includegraphics{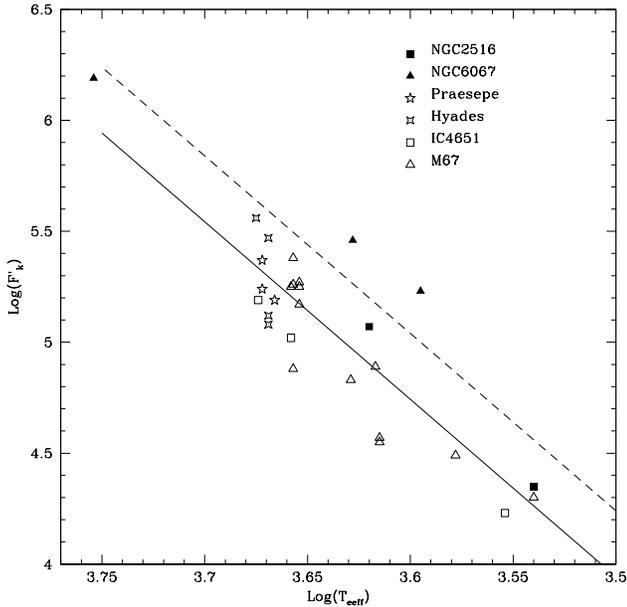}}
\caption{ Log F'$_k$ vs. effective
         temperature for the observed open clusters giants.  Stars
         from different
         cluster are marked with different sysmbols. The M67 data of
         Dupree et al. (1999) are also included in the Figure.}
         \label{Fig4}%
    \end{figure}

Figure 4 confirms the well known trend  between stellar activity and effective 
temperature 
in evolved stars: the higher  the temperature, 
the higher the chromospheric emission. 
 In addition it can be noticed that:
\begin{enumerate}
\item The Hyades giants show a spread of chromospheric fluxes of up to 0.4 dex. 
Since this 
difference is known to be largely caused by 
stellar cycles (Baliunas and Vaughan 1985, see also Stern et al. 1995 
for a discussion on the 
X-ray emission), this first result provides an estimate of 
the cosmic spread of 
chromospheric activity in evolved stars: 
a difference of a few dex can be found even in  otherwise almost
identical  stars.

\item The Hyades and Praesepe giants, which have similar ages, 
temperatures  and masses, also show the same mean level of
chromospheric activity. This is reassuring, because it indicates that 
it may  indeed be possible to represent (within the cosmic spread of
point 1) chromospheric activity as a function of 
macroscopic stellar parameters.  The fact that the chromospheric 
levels are the same among Praesepe and Hyades giants  is also 
in very good agreement with the finding by Randich and Schmitt (1995) 
of a similar  X-ray luminosity among  the
evolved stars of these two clusters.

\item The lower mass RGB stars (IC4651 and M67) form a well defined, 
{\it narrow} sequence with emission slightly lower than the
Hyades and Praesepe clump giants, while the M67 clump stars (located at 
$\log T_{eff}\sim 3.66$) have  
fluxes similar to  the younger 
Hyades and Praesepe.  

\item It is  well known that when field stars are plotted 
in the same activity-temperature diagram, they show a large spread. 
In Figure 4 we may observe how, introducing the $\sim$ 4.3 M$_{\odot}$ 
stars from  NGC 2516 and 6067, the spread is enhanced, in that the 
more massive 
stars tend to show 
higher levels of chromospheric activity for a given T$_{eff}$ than the lower 
mass stars. 
In the same Figure the F'$_k$ $\propto$ T$_{eff}^8 \times M $ 
relationship found by 
PB92 is also shown. The continuous line represents 
what is expected for the Hyades and Praesepe clusters according 
to this law: the data are well 
reproduced.  In addition the broken line gives the predictions 
for 4 M$_{\odot}$ stars. Clearly, the high mass clusters  confirm these 
predictions.
\end{enumerate}

\subsection{Field Stars}

The H-R diagram of the field stars observed is given in Figure 5. 
In this diagram the evolutionary tracks for several  masses (1, 1.3, 1.6 2, 3, 
and 5 M$_{\odot}$ ) are also shown, together with the observed
cluster stars (including the M67 stars from Dupree et al. 1999).
To  make the Figure more readable, cluster stars are grouped in 3 symbols, 
according to the  
cluster turn-off masses:
stars of  NGC 2516 and 6067 as  filled pentagons, stars of Hyades and Praesepe 
as filled squares, 
stars from M67 and IC4651 as filled triangles.  
In Figure 5 the continuous lines refer to the RGB, while the broken lines 
represent the 
1 and 1.6 M$_{\odot}$ HB. 

Clearly the field stars span a large range in  mass and evolutionary status. 
We also note  the good agreement between the cluster turn-off 
masses as given in Table 2 and those indicated by the evolutionary tracks: 
differences can be ascribed 
to the photometry adopted, set of tracks used and to differences in the 
effective temperature scales adopted. We emphasize that a detailed study of 
the colour magnitude diagram of these  
clusters is {\it beyond the scope  of the present work}, and that for our
purposes the agreement shown in Figure 5 is excellent. As expected, 
there is some degeneracy in allocating masses in the region of the 
yellow giants, where He-burning Hyades-like and M67-clump like stars tend to 
group. 
The spread of stars in this region in fact reflects the age dispersion (and to a 
lower
extent the metallicity dispersion) among the 
giants in the solar neighborhood (Girardi et al. 1998). 

   \begin{figure*}
\resizebox{\hsize}{!}{\includegraphics{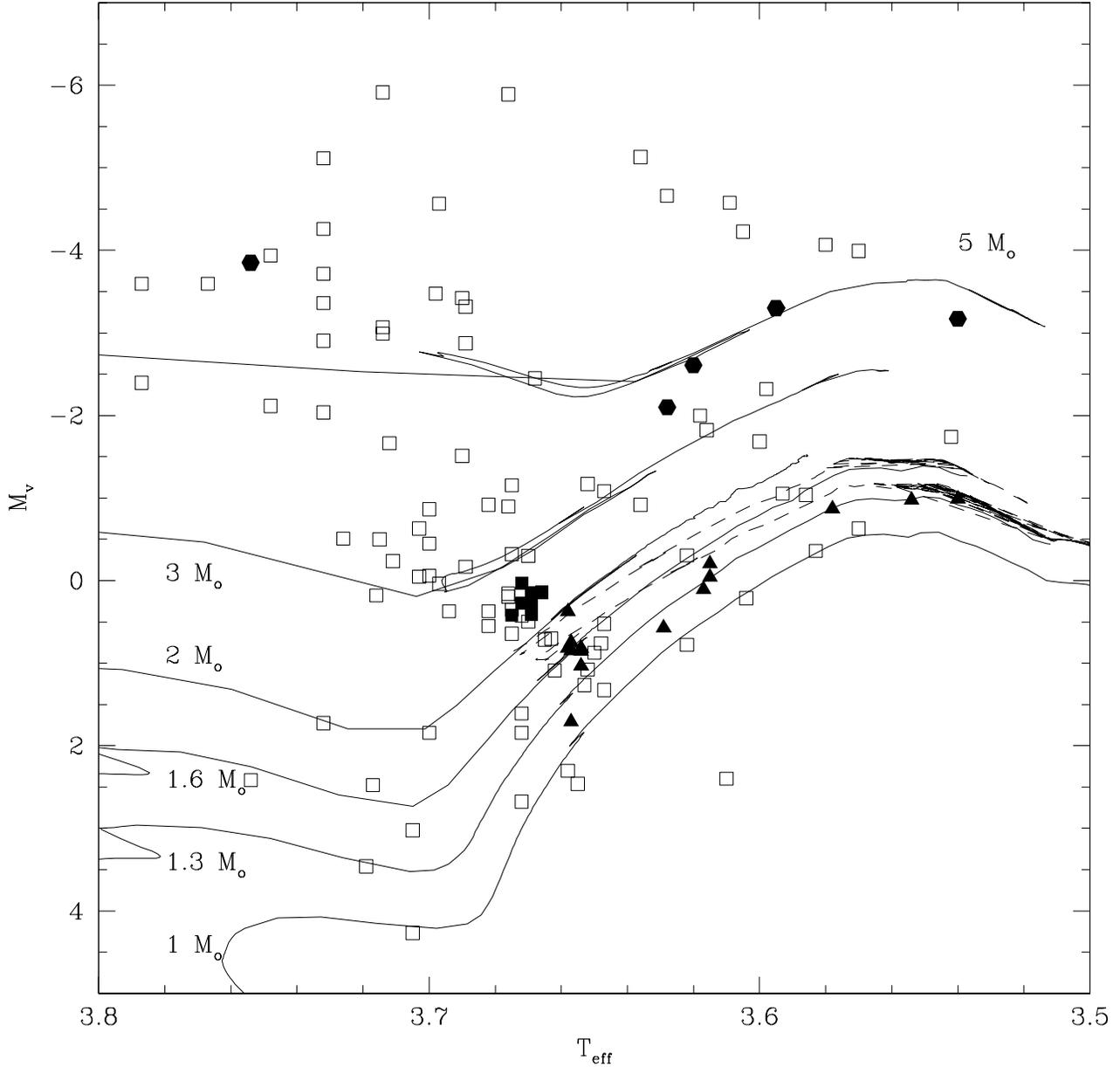}}
\caption{H-R diagram for the sample
         stars. Cluster stars have different symbols, depending on the
         cluster turnoff mass: filled triangles IC4651 and M67 stars; 
          filled squares: Hyades and Praesepe; filled pentagons:
         NGC2516 and 6067 stars; open squares: field stars}  
       \label{Fig5}%
    \end{figure*}

In Figure 5 we only plot those stars of Table 3 with chromospheric emission 
determination. From Table 3 it  emerges clearly 
that the  stars with no detected 
chromospheric emission
are  confined to two well determined groups: 
class Ia supergiants and stars with (V-R) smaller than $\sim$ 0.4. 
It is tempting to say that these stars 
define regions where chromospheric activity is not present, for instance  
because they are so hot
that they 
do not possess a subphotospheric convective zone. However this cannot be 
determined
uniquely on the basis of the Ca II data, because several causes may make the 
detection of Ca II 
emission impossible in these stars: 
\begin{itemize}
\item For the hot stars  high rotational velocities  will tend to smear out the 
Ca II core, and  make the detection of the Ca II minima impossible; in addition
high photospheric contribution at 3933 {\AA} will also  
 make the contrast in the Ca II core very shallow.

\item For the three Ia supergiants, we know that these stars have large line 
broadening, 
with atmospheres not in static 
equilibrium. This line broadening may wipe out the Ca II core, making its 
detection 
impossible. It would not be surprising, however, if the atmospheres 
(or chromospheres, if present at all) of these stars would be very different 
from those of the other late-type giants forming the sample (see e.g. Achmad et 
al. 
1997). 

\end{itemize}

In Figures 6a,b  chromospheric fluxes are shown for all field stars as a 
function of 
effective temperature and measured rotational velocity, respectively. Clearly a 
strong dependence 
of chromospheric activity on both quantities is present, although with a 
large scatter.

In Figure 6b two active stars have low rotational velocities. 
Since our sample contains  $\sim$ 100 stars, some of the 
objects are expected to have very high inclination angles, 
and therefore these high activity,
low {\it Vsini}  stars are likely to be pole-on candidates.
 
   \begin{figure*}
\resizebox{\hsize}{!}{\includegraphics{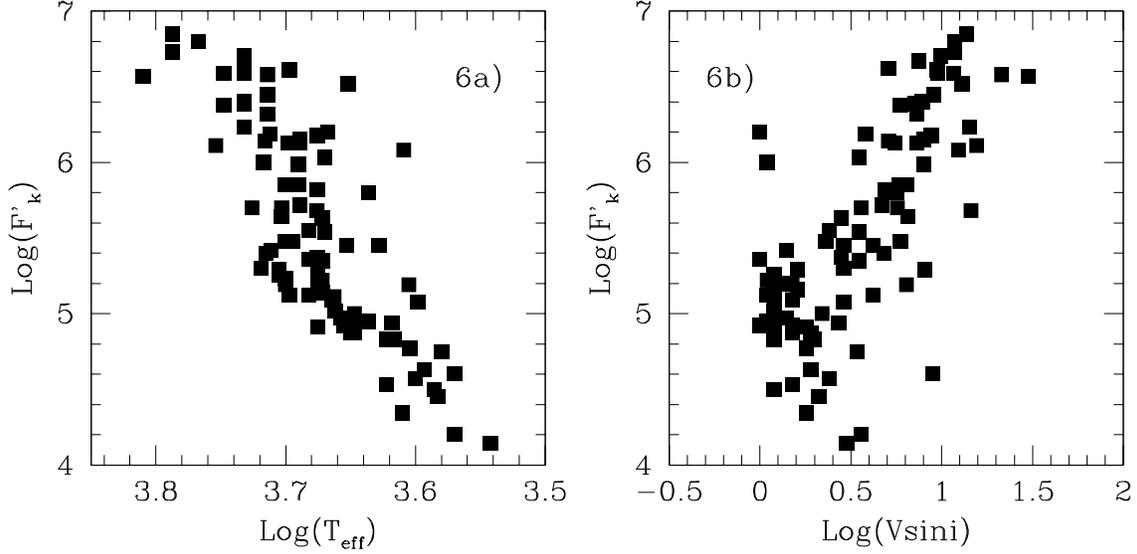}}
 \caption{ Figure 6a: Log F'$_k$ vs T$_{eff}$ 
for the sample stars; Figure 6b: Log F'$_k$ vs {\it Vsini} }
       \label{Fig6}%
    \end{figure*}

   \begin{figure}
\resizebox{\hsize}{!}{\includegraphics{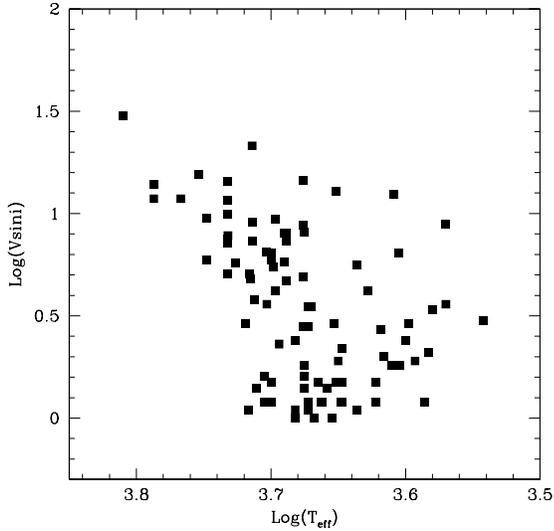}}
\caption{{\it Vsini} 
 vs T$_{eff}$ for the sample stars. }
       \label{Fig7}%
    \end{figure}

{\it Vsini} and T$_{eff}$ are on the other hand only slightly
 correlated in our sample, 
as shown in Figure 7;  this allows us to  fit  activity with both quantities,
obtaining: $$ F'_k \propto  T_{eff} ^{\alpha} {\it Vsini} ^{\beta} $$
with $\alpha$ = 8.01 $\pm$ 0.7 and $\beta$ = 0.8 $\pm$0.1 and a standard 
deviation of 0.32 dex.
Eliminating from the sample the two pole-on candidates of Figure 6b
the fit changes only marginally ($\alpha$ = 7.7 and $\beta$ = 0.9) with a 
standard deviation of 0.29
dex. When considering the intrinsic chromospheric variability and all the 
uncertainties
involved in the analysis (including the {\it sini}), this can be considered a 
good representation
of the data set.
A similar dependence was found by Strassmeier et al. (1994) by analyzing a 
sample containing 
normal and active giants. 

When comparing this result with that of PB92, we find  the same strong 
dependence of
activity on T$_{eff}$, and a similar scatter in the final relationship.
The fact that PB92 found a linear dependence of chromospheric activity 
on stellar mass and now  a similar
dependence on {\it Vsini} is found, 
suggests that  {\it Vsini} should scale with the stellar 
mass: for the same effective temperature, more massive giants 
should  have higher rotational velocities. 

One immediate explanation for the observed correlation is to assume that  
the dependence on T$_{eff}$ and {\it Vsini} represents two well separated 
mechanisms: 
the dependence on T$_{eff}$ being a  component related to the stellar 
structure, 
and the {\it Vsini} term representing an additional source of chromospheric 
heating, likely
of  magnetic origin. 
This scheme has been proposed in the past by several authors (see e.g.
Rutten and Pylyser 1988 and references therein), and it is natural to 
associate the 'temperature' 
component with that generated by acoustic heating (see e.g. Cuntz et al. 1997).
The acoustic theory has been developed  
in the last 20 years, and some of the most recent  
works (Buchholz et al. 1998) predict an activity-temperature 
dependence  
very close to the one observed. 
It is not clear to the authors if, for instance, any dependence of acoustic 
heating  should be expected with stellar mass, due to the different internal 
structure 
of the stars sharing the same effective temperature. 

The presence of stellar cycles in the Hyades giants as well as the 
measurements of Ca II variability induced by rotational modulation could 
on the other hand 
question the relevance of the acoustic heating mechanism and these 
observed phenomena  suggest that 
acoustic heating  may represent only  a small fraction of the 
overall chromospheric heating budget.
In the specific case of the Hyades giants for instance, 
the differences observed 
would indicate that the magnetic part is at least 4 times higher than the 
acoustic contribution. 
The data of the Hyades giants, however tell us also 
something else: that even if a correlation between 
activity and {\it Vsini} is found, a simple correlation 
can  only have a limited range of 
accuracy: indeed among the Hyades giants, the difference in measured 
{\it Vsini} is rather small (cfr. Table 2), and their 
difference in activity cannot likely be ascribed directly to a  
difference in stellar rotation. This indicates that any activity - rotation
relationship, while it can provide with  a broad agreement with the 
observations, it may hide rather  complex physical mechanisms, 
 possibly similar to those in force in the sun.

A different approach considers that
in the presence of a composite sample, we can divide the sample according 
to the stellar masses and  follow the evolution of the 
chromospheric activity along each of the evolutionary tracks. 
For instance it is clear that while a star evolves along its track, 
its rotational velocity will also evolve, giving 
a dependence of rotational velocity from effective temperature 
and the plot of figure 7 could therefore be the result of 
merging many stars with different masses and evolutionary histories in 
the same diagram.  

In the following we will  
analyze  the data by studying the  possible
relationship between activity and 
angular momentum evolution, making use of the evolutionary tracks. 
A somewhat similar approach was adopted by Rutten and Pylyser (1988); these 
authors found that the post main sequence evolution of  angular momentum and of
 stellar activity could be, at least qualitatively, explained 
in terms of angular momentum conservation for the most massive stars, 
while for stars of 2 M$_{\odot}$ or lower, this law would predict 
a much too high activity (and rotation rate) for stars having
intermediate temperatures.
The advantage of using Hipparcos based parallaxes, 
homogeneously determined chromospheric fluxes and 
stellar rotational velocities, together with the information coming from the
clusters, should provide new hints on this topic. 
 
We have therefore subdivided our sample according to the evolutionary tracks: 
if stars are closer than 0.01 in Log(T$_{eff}$) and 0.3 in M$_v$ 
to the 1, 1.3, 1.6, and 2 M$_{\odot}$ evolutionary tracks,
they were given the respective  masses. For more massive stars the M$_v$ 
requirements were relaxed: 
for 3 M$_{\odot}$ stars all objects within M$_v$ (-0.3 + 0.7) from the track 
have been  given 3 M$_{\odot}$, while all stars 
above the 5 M$_{\odot}$ curve and 0.7 
magnitudes below it were considered as 5 M$_{\odot}$
stars.  
Clearly this division suffers from some 
limitations: for example several stars  will be common 
to two subsamples; 
in addition, since in this subdivision we have considered only the RGB, 
in the region of the yellow giants we will attribute higher masses 
to some clump low mass stars; finally, the 5 M$_{\odot}$ subsample will be 
more composite than the others. 

\subsection{Empirical relationships for different masses}

Under the hypothesis  that all stars belonging to a  subsample represent 
different evolutionary stages of objects having otherwise 
similar characteristics, 
linear fits of Log F'$_k$  vs different quantities performed 
for the different subsamples will tell us how chromospheric activity 
will evolve for the stars of a given age (or turnoff mass). 
We have performed fits of the following 
form:  F'$_k$ $\propto$ X$^{\alpha}$, 
where X can be either the rotational velocity {\it Vsini}, the 
angular velocity $\Omega$ or the effective temperature T$_{eff}$. 
The results of the fits are given in Table 4, together with 
the dispersions. 

We can immediately 
notice that, when subdividing the stars in mass ranges, for  the 1.3-3 
M$_{\odot}$ samples, the Ca II emission can rather well be 
represented as a function of {\it Vsini}, 
without any need for an additional  parameter, with a dependence
of activity on {\it Vsini} close to 1. 
For the 1 M$_{\odot}$ Ca II does not depend on rotational velocity. The 1
M$_{\odot}$ case is very interesting because, as we will see later, 
taking the data at face value, a rather peculiar angular momentum evolution 
is required. For the 5 M$_{\odot}$, while stellar activity depends
strongly on T$_{eff}$, it does not depend on {\it Vsini}. 

Since {\it Vsini} is a quantity which involves the stellar radius, 
rotational periods could be a more appropriate parameter to represent 
any rotationally induced effect. We have  computed for each star the 
stellar radius from the data in Tables 1,2,3 and derived periods. 
The results of the fits are also given in Table 4 and the diagrams are 
shown in Figure 8. 
With the exception of the 5 M$_{\odot}$ stars, very good fits can be obtained
by using this parameter. 
Finally, good fits for all our subsamples can be obtained by using 
T$_{eff}$ as governing parameter. 

   \begin{figure*}
\resizebox{\hsize}{!}{\includegraphics{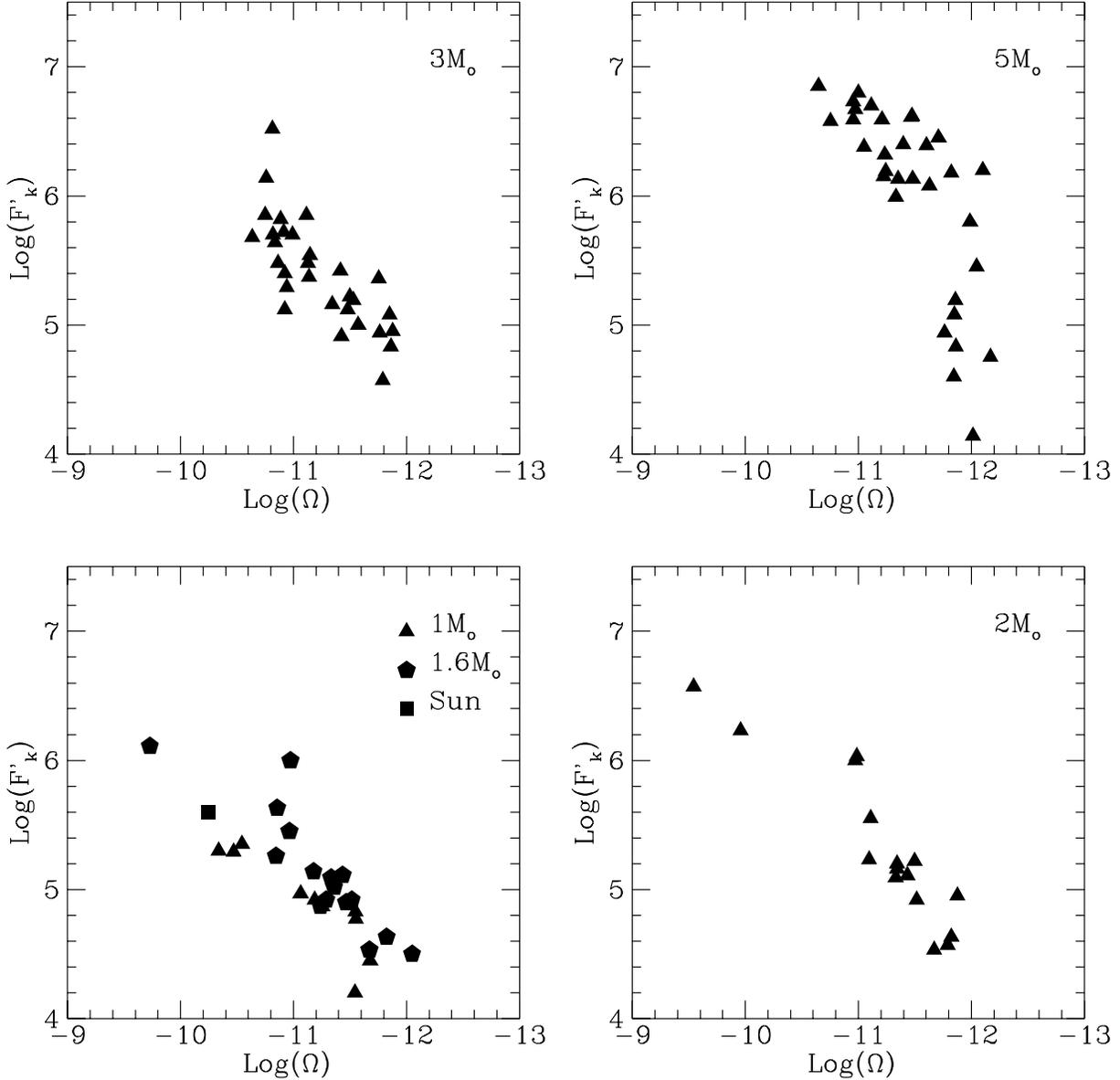}}
  \caption{Log F'$_k$ vs. $\Omega$ for the different 
mass subsamples. In the first panel both 1 and 1.6 M$_{\odot}$ stars are 
given, with different symbols. The sun is also plotted as a filled square }
       \label{Fig8}%
    \end{figure*}

\setcounter{table}{3}
   \begin{table*}
      \caption[]{Empirical relationships Log F'$_k$ found when analyzing
      separately the
      sample stars with different masses. }
         \label{table4}
         \begin{center}
	\begin{tabular}{l|r|r|r|r|r|r|r|r|r|r}\hline
\multicolumn{1}{c}{Function}  &  
\multicolumn{1}{c}{ $\alpha$ 1M$_{\odot}$} & 
\multicolumn{1}{c}{$\sigma$ 1M$_{\odot}$} &
\multicolumn{1}{c}{ $\alpha$ 1.6M$_{\odot}$} & 
\multicolumn{1}{c}{$\sigma$ 1.6M$_{\odot}$} & 
\multicolumn{1}{c}{ $\alpha$ 2M$_{\odot}$} & 
\multicolumn{1}{c}{$\sigma$ 2M$_{\odot}$} & 
\multicolumn{1}{c}{ $\alpha$ 3M$_{\odot}$} & 
\multicolumn{1}{c}{$\sigma$ 3M$_{\odot}$} & 
\multicolumn{1}{c}{$\alpha$ 5M$_{\odot}$} & 
\multicolumn{1}{c}{$\sigma$ 5M$_{\odot}$} \\ 
                       \hline
Vsini    &0.0 &0.37 & 0.92 &0.38 &1.03 & 0.42 & 0.87 & 0.31 & 1.56 & 0.59 \\
$\Omega$ & 0.65 & 0.18 &0.70 &0.24 & 0.86 & 0.23 &0.89 & 0.26 & 1.3 & 0.48  \\
T$_{eff}$& 7.5  & 0.15  & 9.99 & 0.2 &10.35  & 0.26 & 6.76 & 0.34 & 10.5 & 0.27 
\\
             \hline
         \end{tabular}
	\end{center}
    \end{table*}

From Table 4 several points emerge: 

a) That the representation of the chromospheric flux 
is better obtained as a function of $\Omega$ than {\it Vsini}

b) That the Ca II fluxes can be well represented as a function of the 
angular velocity only, but 
for stars of $\sim$ 5 M$_{\odot}$ and above the fit is rather 
poor. A large slope is required for these stars. A clear drop in the 
relationship is clearly present in Figure 8 when the coolest supergiants are 
considered: these stars show {\it higher} rotational velocities 
(or lower Ca II fluxes) than expected from extrapolating the trend 
given by the hotter stars.

c) While the dependence on T$_{eff}$ can be acceptable for most 
mass intervals, it may be difficult to explain how this dependence 
should strongly change in slope among the different subsamples if T$_{eff}$ 
was the only relevant parameter.

\subsection{Angular Momentum Evolution}
The measured {\it Vsini} and the attribution of masses allow also 
a preliminary analysis of the evolution of the stellar angular momentum. 
We have computed from the Padua models the momentum of inertia I of 
the stars along the evolutionary tracks.  For spherically symmetric 
distribution 
of density $\rho(r)$ I is given by:
\begin{equation}
\frac{8\pi}{3} \int_0^{R} \rho(r) r^4 dr
\end{equation}
where $r$ is the radial coordinate and R the stellar radius. 
We recall that the adopted models do not include mass losses, angular 
momentum redistribution from the interior or braking.
In the following we will not perform fits to the observed {\it Vsini}, 
rather  we will verify if the observed {\it Vsini}  distribution 
can be reproduced in terms of simple scaling laws of the form 
I$\Omega ^{\beta} = Const $.

\subsubsection{Low Mass stars}

   \begin{figure*}
\resizebox{\hsize}{!}{\includegraphics{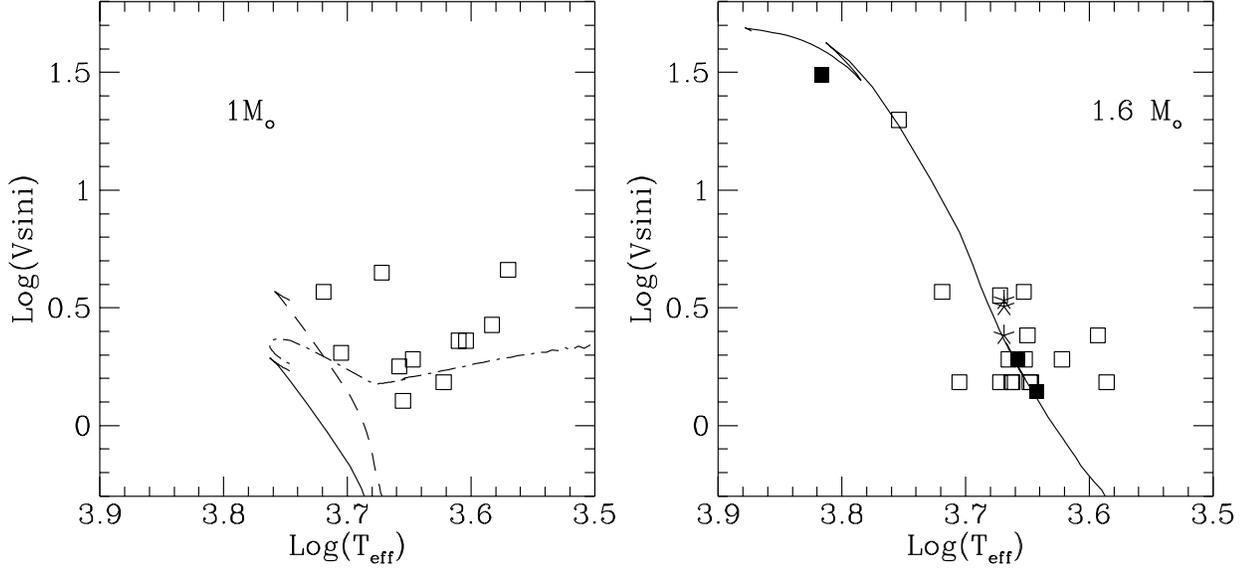}}
 \caption{{\it Vsini} vs. T$_{eff}$ for the 
low mass subsamples.  Figure 9a: for 1M$_{\odot}$ stars. The continuous line
represents what is expected in the case  I$\Omega$=Const. The dashed-dotted line
represents what is expected if a I$\Omega ^2$=Const law would hold. Figure 9b: Same as 9a),
but for 1.6 M$_{\odot}$ stars. Rotational velocities derived from 
period measurements (Choi et al. 1995) are given as starred; the filled dots 
represent the mean rotational velocities measured in the NGC 3680 stars: 
8 (single) turn-off and 4 giants. }  
       \label{Fig9}%
    \end{figure*}

Figure 9a,b show the rotational velocities  
for the 1 and 1.6 M$_{\odot}$ stars respectively. 
The observed {\it Vsini} of table 1 and 2 have been multiplied by 4/$\pi$ 
to take into account the projectional effects. 

The first striking point is that, if taken at face values, the {\it Vsini} 
of the 1 M$_{\odot}$ stars is independent of the stellar effective 
temperature and therefore of their evolutionary status. 
This is at odds with the 
intuitive picture, from which we would expect that the stars slow down 
rather quickly after they leave the main sequence and that further 
deceleration could happen if magnetic braking is at work. 
In Figures 9a,b the lines represent what is expected from the simplest 
hypothesis: 
$$I\Omega = Const(M) $$

or V = V$_o \times 1/I $

The 1 M$_{\odot}$ curve has been adjusted to reproduce the solar equatorial 
velocity of 2 km/sec (continuous line in Figure 9a), 
while for the 1.6 M$_{\odot}$
stars the curve has been adjusted to reproduce the mean distribution of the 
giants. 

Clearly, this simple law does not work for the 1 M$_{\odot}$ stars: 
giants rotate too fast to obey the angular momentum conservation law. 
Of course one could consider that the solar zero point is wrong and that 
a typical solar type star has a rotational velocity as large as twice
that of the Sun; but even when considering this case (broken line in Figure 
9a) the conservation of angular momentum would predict far too low 
rotational velocities for the giants. Only with a law of the type 
I$\Omega ^2$=Const  can the data points  be reproduced, as is shown by the 
dashed dotted line in Figure 9a, which represents the behavior of this law. 
It is interesting 
to observe, however, that in case activity and rotational velocities were 
related, these relatively high {\it Vsini}  for the low mass stars would 
explain the relatively high level of activity observed among the
globular clusters giants (Dupree et al. 1990, 1994). 
But we have to stress that  the {\it Vsini} values measured in these 
stars are typically of a few km/sec, therefore  measurement uncertainties 
(systematics in particular) could  in this regime play a crucial role. 

   \begin{figure*}
\resizebox{\hsize}{!}{\includegraphics{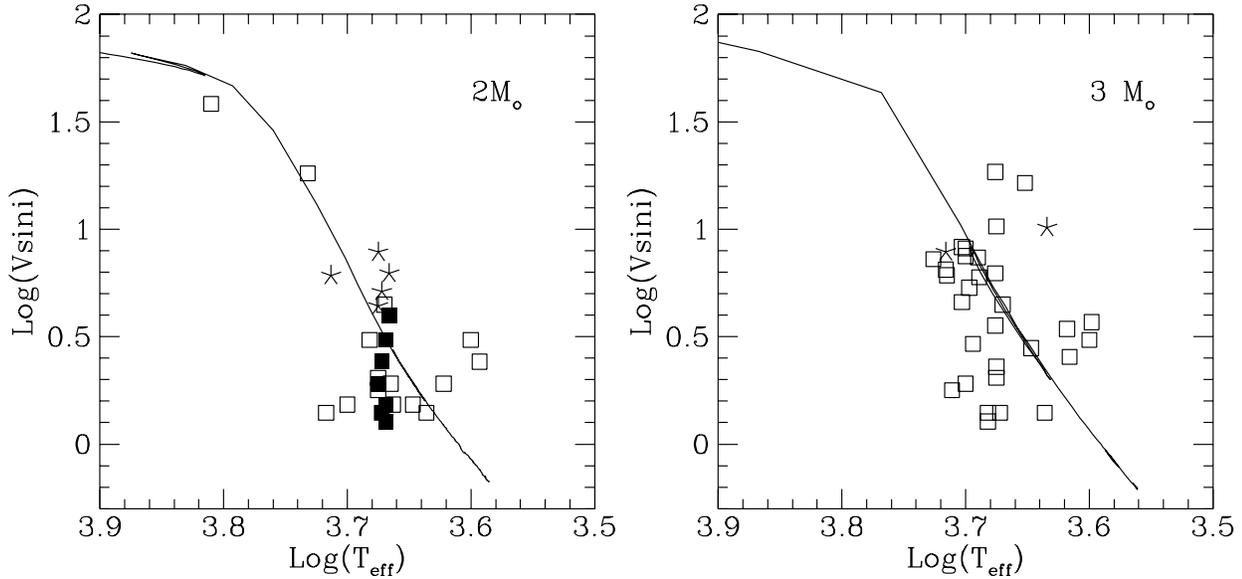}}
\caption{ {\it Vsini} vs. T$_{eff}$ for the 
intermediate  mass subsamples.  Figure 10a: for 2M$_{\odot}$ stars. 
The continuous line
represents what is expected in the case  I$\Omega$=Const. 
 Rotational velocities derived from 
period measurements (Choi et al. ) are given as starred; the filled dots 
represent the mean rotational velocities measured in the NGC 3680 stars: 
8 (single) turn-off and 4 giants and the values for the Hyades and Praesepe 
stars of Table 2. Figure 10b: the same for the 3 M$_{\odot}$ stars } 
       \label{Fig10}%
    \end{figure*}

Pending more and more accurate data on 1 M$_{\odot}$ giants, we can
conclude that  the angular momentum evolution of these 
low mass stars does not follow a I$\Omega$=Const law, but possibly a 
more steep one. 
From the physical point of view it is difficult to explain this behavior: 
possibly large mass losses in the RGB phase (which are also required 
for the interpretation of the color-magnitude diagrams of globular clusters)
with winds not suffering magnetic backing could explain part of this 
behavior, but we would expect these mass losses, if at all, to act mostly 
towards the coolest part of the RGB only. 

For the 1.6 M$_{\odot}$ stars the situation is different: 
the conservation of angular momentum law reproduces very well most stars. 
To better look into this point, in Figure 9b the rotational velocities
of three 1.3-1.6 M$_{\odot}$ stars having measured rotational periods 
are shown as starred points (Choi et al. 1995). Finally, in the same Figure 
the filled squares represent the CORAVEL {\it Vsini} of stars belonging to the 
open cluster NGC 3680 (mean of 8 turn-off single stars and mean of 4 single 
giants, corrected by 4/$\pi$, Nordstrom et al. 1996,1997): the agreement 
between the  prediction of the constant angular momentum law derived for the 
field stars and the NGC 3680 points is impressive. 

It is worth recalling that there is a basic structural difference 
between the 1 and the 1.6 M$_{\odot}$ stars of figure 9a and 9b: 
the low mass stars present a radiative nucleus and a convective envelope 
on the main sequence, whereas the opposite situation is found 
in the more massive stars. This may also produce relevant 
differences in the evolution of the stellar rotational velocity.

\subsubsection{ Intermediate mass stars (2-3 M$_{\odot}$)} 

Figures 10 a,b show the 2 and 3 M$_{\odot}$ subsamples with, superimposed, 
the results obtained by assuming angular momentum conservation. 
The agreement is very good. 
As in Figure 9b starred points are {\it V} as derived from the  Choi et al. 
(1995) periods. The filled squares represent the Hyades and Praesepe 
measured {\it Vsini}.  It is worth noticing that out of the 
Choi et al. stars, 3 are belong to Praesepe and 1 to the Hyades; 
for these stars we can therefore compare the
{\it Vsini} with the {\it V} derived from the  observed periods. {\it Vsini} 
values are systematically lower than the {\it V} derived from the
periods.  
The Hyades and Praesepe giants have a mean {\it Vsini} 
of 1.74 km/sec, while for the 4 Praesepe and Hyades giants from Choi
the mean V is of 5.4 km/sec.
This result indicate that the Coravel {\it Vsini} may 
slightly underestimate the {\it Vsini} for these stars, in line with
the offset between Coravel and Gray measurements quoted above. 

The angular momentum conservation law would predict a V$\sim$ 65 km/sec 
for the Hyades turn-off stars, which is comparable to, but lower than the 
observed mean turn-off velocity of 100$\pm$50 km/sec (Gaige 1993). 
Although we cannot consider this discrepancy as significant (in fact
the Hyades turnoff value would be perfectly reproduced if a systematic
effect is
present and the line was passing through the Choi et al. (1995) points
of Figure 10a), still the difference could be due to the 
simplification adopted in our  model; 
a factor two slow down between turn-off and giant stars  is predicted
for instance by models  
adopting internal angular momentum redistribution (Endal and Sofia 1979).

\subsubsection{ High mass stars (5M$_{\odot}$)} 

Figure 11 shows the behavior of the rotational velocity vs. effective 
temperature
for the 5 M$_{\odot}$ stars. Obviously the I$\Omega$=Const law (continuous
line) does not represent the data. As for the 1 M$_{\odot}$ stars 
the distribution of rotational velocities 
requires a much steeper dependence on the angular velocity. In the same Figure
the I$\Omega ^2$ = Const law is given, which approximates much better the 
observations. 
As found for the Ca II analysis in the previous chapter, the cool 
supergiants seem to rotate too fast with respect to what is predicted 
by their hotter 'precursors'. We do not have any explanation for this 
phenomenon, we caution only that for these stars the nominal 
uncertainty in {\it Vsini } is likely underestimated, since in 
these stars other line broadening mechanisms may be comparable or superior 
to those induced by {\it Vsini }.

   \begin{figure}
\resizebox{\hsize}{!}{\includegraphics{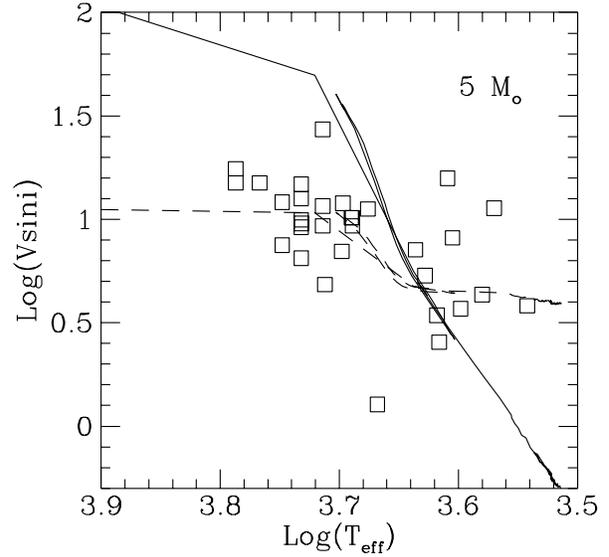}}
\caption{{\it Vsini} vs. T$_{eff}$ for the 5M$_{\odot}$ stars. The 
continuous line represents what is expected in the case  I$\Omega$=Const. 
The dashed-dotted line represents what is expected if a I$\Omega ^2$=Const 
law would hold. }
       \label{Fig11}%
    \end{figure}

\subsection{ Consistency}

Before deriving any conclusion, we believe that it  is important to 
perform a cross check of the previous analysis. 
We shall in this section outline a summary of the 
analysis previously performed. 

We have first seen how  F'$_k$  can be well
fitted as a function of $\Omega ^{\alpha}$, with $\Omega$ derived 
from {\it Vsini}, M$_{Bol}$ and T$_{eff}$ as given  in tables 1,2,3. 
As a second step we have investigated the dependence of the observed 
{\it Vsini} on  the I$\omega ^{\beta}$ = Const scaling law 
by using the momentum of inertia I as computed by theoretical 
models, finding that the data are consistent with $\beta$=2 for the 1 and 
the 5 M$_{\odot}$ stars, and $\beta$=1 for the 1.6-3 M$_{\odot}$ stars.
Combining these results we  expect that for each 
subsample the derived Ca II chromospheric fluxes should depend 
on the momentum of inertia I (as computed by the models) according to 
F'$_k \propto I^{-{\alpha}/{\beta}}$, where $\alpha$ for the different 
subsamples is  given 
in Table 4 and $\beta$=1 or 2 depending on the stellar mass subsamples.

   \begin{figure*}
%
%
\resizebox{0.80\hsize}{!}{\includegraphics{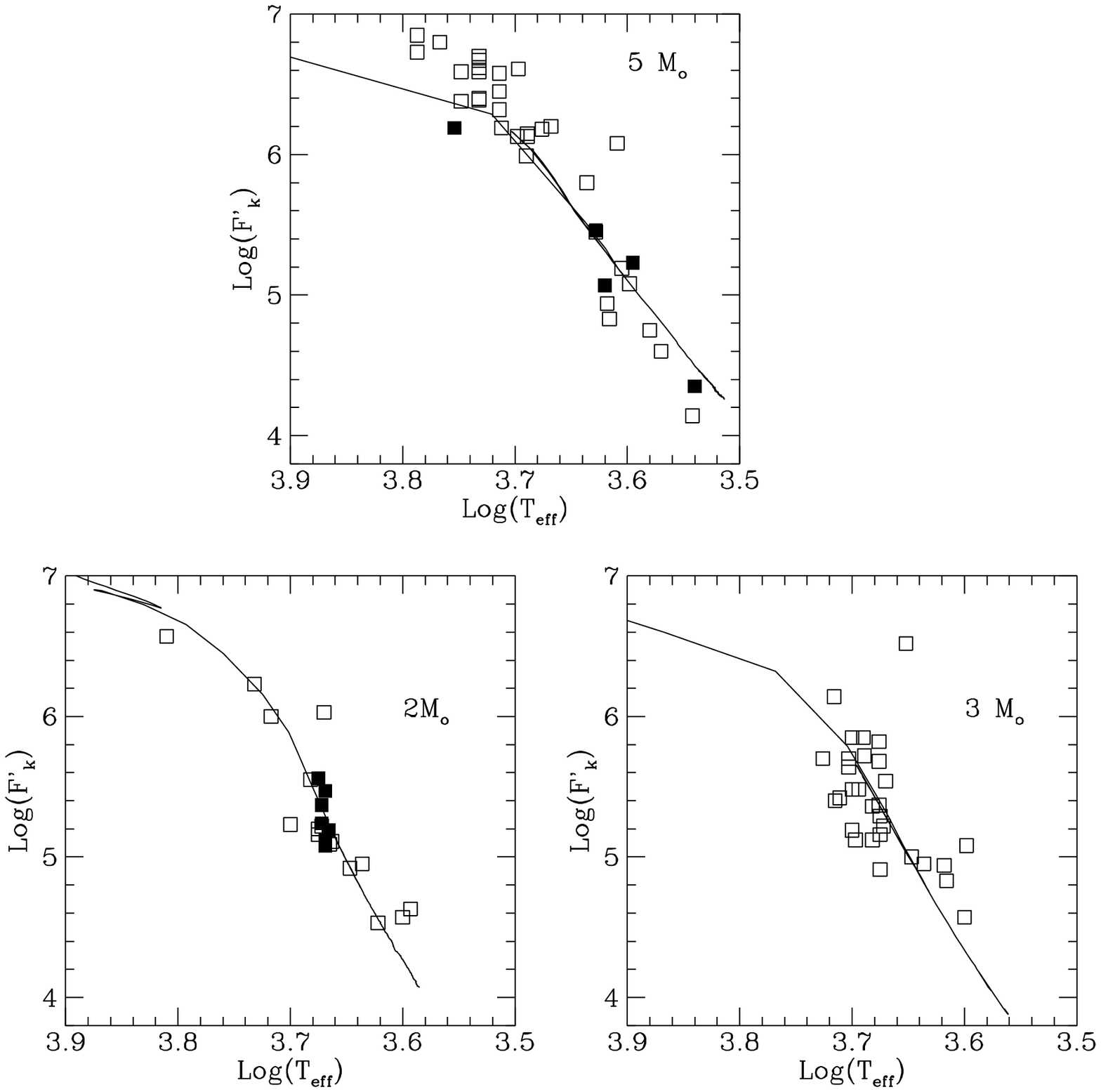}}
\resizebox{0.80\hsize}{!}{\includegraphics{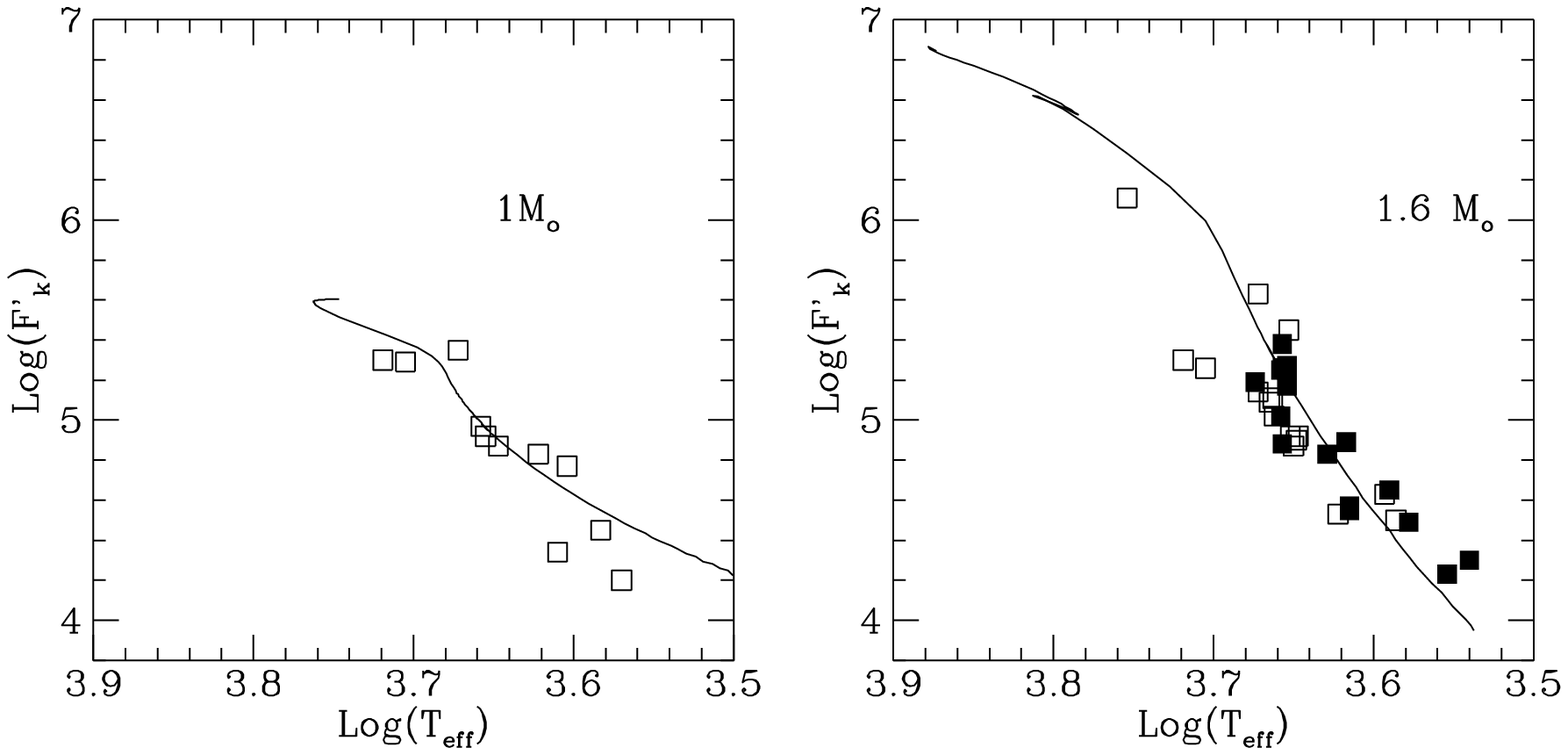}}
\caption{ 
Verification of the results obtained: the 
Ca II fluxes Log F'$_k$ 
vs T$_{eff}$  diagrams are reproduced by the expected log F'$_k \propto 
I^{-\alpha / \beta}$  with $\alpha$ as derived from Table 4 and $\beta$= 1 for 
1.6-3 M$_{\odot}$ and $\beta$=2 for 1 and 5 M$_{\odot}$ (cfr. Figures 9,10,11)
}
       \label{Fig12a}%
       \label{Fig12b}%
    \end{figure*}

We have therefore used the models and the 
previously determined exponents to verify this consistency and the results are
shown in Figures 12a-e for the different mass subsamples.  
As can readily be seen, the results are very satisfactory, 
confirming therefore the consistency and the quality of our analysis. 
We find the results in Figure 12a-e very encouraging, although 
several improvements could be performed in future studies. 

This analysis holds for our sample  stars, 
whose angular momentum behavior is not dominated 
by those phenomena, like tidal locking, which 
determine the dynamic behavior in short period binaries. For those cases 
rotational periods should instead be taken explicitly into
account. 

\section{Conclusions}

By providing a large, {\it homogeneous}  sample of Ca II chromospheric fluxes 
{\it and} rotational velocities 
for field {\it and } cluster F-K giants, thanks to the use of {\it Hipparcos} 
accurate parallaxes 
and {\it evolutionary tracks}, we have found first approximation 
laws which govern  chromospheric activity and angular momentum evolution 
in F-K giants and supergiants. In particular:

\begin{enumerate}

\item Hot giants and Ia supergiants do not show Ca II emission. 
We cannot conclude from the Ca II only if this 
is due to the absence of a chromosphere
or to other factors (high {\it Vsini}, large turbulence)

\item When considering young clusters, the results are in agreement 
with  what  is
observed by PB92:  for a given effective temperature 
more massive stars tend to have higher chromospheric fluxes than low
mass stars.

\item Hyades and Praesepe giants show the same {\it mean} 
level of chromospheric activity,  as expected 
for stars having the same mass and effective temperature. 

\item However Hyades stars show a spread of up to 0.5
 dex in chromospheric fluxes. 
This difference shows that stellar cycles produce some 'cosmic' 
scatter among stars which are otherwise very similar. This scatter would  limit 
our prediction capabilities even in the absence of measurement errors.

\item When analyzing the whole sample the Ca II  fluxes 
can be well represented by 
F'$_k$  $\propto$ T$_{eff} ^{7.7} \times {\it Vsini}^{0.9}$ .
\item On  dividing the sample into stars of different masses by using 
evolutionary 
tracks, we find that we can represent F'$_k$ for all subsamples but the 5 
M$_{\odot}$ stars as a function of the angular velocity: 
F'$_k \propto \Omega^{\alpha}$. Also, in each subsample F'$_k$ can be 
well represented as a function of the stellar effective temperature 
T$_{eff}$ (cfr. Table 4)

\item By using the momentum of inertia I as deduced by theoretical models 
we have verified that the observed rotational velocities can be represented 
by simple I$\Omega ^{\beta}$ = Const laws. 1.6-3 M$_{\odot}$ giants 
are consistent with angular momentum conservation law ($\beta$=1), 
while for the 1 and 5 M$_{\odot}$ giants a steeper, $\beta$=2 law seems 
more appropriate. (We caution, however, that for different reasons
these two samples may be heavily affected by systematics errors in
their {\it Vsini} measurements). 

\item The {\it Vsini} data for 1 M$_{\odot}$ stars do not show any 
dependence on the stellar evolutionary status. This result 
is at odds with the intuitive expectations.

\item Finally we have checked the consistency of our analysis 
and we have been able to reproduce the observed Ca II chromospheric 
fluxes F'$_k$ as a function of the theoretical momentum of inertia
using  the results from the above analysis: F'$_k \propto I^{-\alpha / \beta}$.

\end{enumerate}

As far as the mechanism governing stellar activity in giant stars is 
concerned, our analysis shows how, with two macroscopic parameters, it
is possible to determine the level of chromospheric activity
in evolved stars. 
This can be explained in two alternative ways: 

a) Chromospheres are powered by two mechanisms: one dependent 
on the stellar rotational velocity and one on the stellar effective 
temperature. 

b) Evolved stars of the same  mass tend to behave 
   similarly, having similar angular momentum for a given 
position on the evolutionary tracks. The angular momentum 
evolution is however different for stars of different masses 
and angular velocity and mass are the only parameters determining the
level of activity. 

We stress that some of the present uncertainties could be 
dissipated by observing subgiants and giants  in  well populated 
clusters. A full self consistent analysis (e.g. including fitting
  of the C-M diagram) with the theoretical models 
would reveal which of the above interpretations is correct. 
Also, measurements of rotational periods in many more giants and 
supergiants will be fundamental in assessing systematic uncertainties in the 
{\it Vsini} measurements.

\begin{acknowledgements}
A  large part of this work was prepared during the stay of LP at UFRN in Natal. 
This work has been possible thanks to the ESO DGDF and Brazilian agency CNPq
grants. The work of LG was funded by the Alexander von Humboldt 
stiftung. We are grateful to L. Achmad for his help in the data
reduction and  P. Bristow  for a careful reading of the manuscript.

\end{acknowledgements}

\newpage

\setcounter{table}{0}
 \begin{table*}
      \caption[]{PB92 stars. Column 1: HD number; Column 2:
      Spectral Type; Column 3: Binarity flag; 
          Column 4: Effective Temperature;
      Column 5: M$_v$ (Hipparcos); Column 6: M$_{Bol}$; 
      Column 7: Log of Ca II chromospheric fluxes; 
   Column 8: {\it Vsini} CORAVEL; Column 9: Vsini
  (Gray and collaborators); Column 10: Color excess e(V-R)}
         \label{table1}
         \begin{center}
	\begin{tabular}{l|l|r|r|l|l|l|r|r|r}\hline
\multicolumn{1}{c}{HD}  &  
\multicolumn{1}{c}{Sp. Type} & 
\multicolumn{1}{c}{Bin} & 
\multicolumn{1}{c}{T$_{eff}$} & 
\multicolumn{1}{c}{M$_v$} & 
\multicolumn{1}{c}{M$_{Bol}$} & 
\multicolumn{1}{c}{LogF'$_k$} & 
\multicolumn{1}{c}{Vsini} & 
\multicolumn{1}{c}{Vsini(Gr)} & 
\multicolumn{1}{c}{e(V-R)} \\
                       \hline
           
2261  & K0III  &    SB  & 3.647  & 0.52 & 0.03 &  4.92 & 1.2 & 2.1 GP & \\
4128  & K0IIICH-1H & S   & 3.670 &-0.30 &-0.67 & 5.54 & 3.5 & 3.0 G &  \\  
9270  & G7IIIa    &  S   & 3.675 &-1.16  &-1.53 & 5.29 & 8.1  &      &   \\    
11937 & G8IIIbCNIV & S   & 3.717 & 2.48  & 2.24 &  6.00 & 1.1  &     &   \\    
13611 & G6II-IIIC  & SB  & 3.700 &-0.87 & -1.07 &  5.19 & 1.5  &     &   \\  
16522 & G8III      & S   & 3.682 & 0.37 & -0.04 &  5.12 & 1.1  &     &  0.03 \\
17652 & G8IIIb     & S   & 3.675 & 0.64  & 0.27 &  5.20 & 1.4 &     &  0.05  \\
21120 & G6IIIFe-1  & SB  & 3.700 & -0.44 & -0.70 & 5.48  & 5.9 &     &   \\    
27256 & G8II-III   & SB  & 3.689 & -0.17 & -0.35  &5.72  & 4.7 &     &   \\    
29139 & K5III      & S   & 3.570 & -0.63  & -1.73  & 4.20 & 3.6  &    &   \\     
36079 & G5II       & SB? & 3.703 & -0.63 & -0.79 & 5.64 & 6.5  & 5.1 G & \\    
39364 & G8IIIwkCN  & S   & 3.652 & 1.08  & -0.63 & 4.92 & 1.5  &     &    \\    
48329 & G8Ib       & SB  & 3.676 & -5.89  & -6.13 & 6.18 & 8.8  & 7.0 GT& 0.37 
\\
50310 & K0III     & SBO  &3.647  & -1.08   &-1.57  &5.00  &2.2  &2.9 GP & 0.06 
\\
50778 & K4III     & S    &3.583  &-0.36  &-1.36   &4.45  &2.1   &      &      \\   
52497 & G5IIa-Ib  & S    &3.690 &-3.42  &-3.60  &5.99  &8     &   & 0.02 \\
57146 & G2Ib      & S    &3.732 &-3.36 &-3.40  &6.70  &9.9   &    &0.12  \\
57623 & F6II   	  & S    &3.767 &-3.60  &-3.56  &6.80  &11.8   &    & 0.22 \\
59890 & G3Ib      & S    &3.697 &-4.57  &-4.71  &6.61  &9.4   &  & 0.12   \\
61064 & F6III     & SB?  &3.810 &-1.18  &-1.28   &6.57  &30.1  &   &       \\   
62345 & G8IIIa    & S    &3.675 &0.35  &-0.02  &5.16  &1.6    &   &      \\     
63700 & G3Ib      & SB   &3.714 &-5.91 &-6.02  &6.45  &9.1    &    & 0.25 \\
65228 & F7II      & S   & 3.787 &-2.34 &-2.33  & 6.85  &13.8    &     &  0.14 \\
67594 & G2Ib      & S   & 3.732 &-5.12  &-5.16  &6.39   &7.2    &     & 0.15  \\
68752 & G5II      & S   & 3.689 & -3.32   & -3.50 & 6.13 & 7.3   &  & 0.09  \\  
71129 & K3III+B2V & S   & 3.609 &-4.58  &-5.23  & 6.08  & 12.4  &   &     \\     
74395 & G1Ib       & S   & 3.732 & -2.04 & -2.08 & 6.67  & 7.5   &    & 0.07 \\
77258 & G8K-1III+A & SBO & 3.670 & 0.50 & 0.12  &  6.03 & 3.5   &    &     \\  
81797 & K3II-III   & S    & 3.600 & -1.69 & -2.39 &  4.57 & 2.4   &    &    \\    
84441 & G1II      & S    & 3.748 & -2.12  & -2.12 &  6.38 & 5.9    &   & 0.14 \\
89485 & G7IIICN-I & S    & 3.682 & -0.92  & -1.12 &  5.36 & 1    & 2.7 G &   \\
92449 & G2-3Ib    & S    & 3.732 & -3.72  & -3.76 & 6.62 & 5.1 &    & 0.17 \\
93497 & G5III+G2V & S    & 3.700 & -0.06 & -0.30 & 5.85 & 6.4 &  6.1 GP & \\ 
98430 & G8III-IV  & SB   & 3.675 & -0.32 & -0.67  & 4.91 & 1.8  &       &  \\    
100407& G7III     &  S   & 3.682 & 0.55  & 0.20 & 5.55 & 2.4  &       &  \\   
102350& G5Ib-II   & S    & 3.690 & -1.51  & -1.81  & 5.85  & 5.8  &      & \\   
105707& K2.5IIIa  & S    & 3.616 & -1.82  & -2.50 &  4.83 &  2  & 2.6 GP & \\  
109379 &G5II      & S    & 3.726 & -0.51  & -0.66 & 5.70  & 5.7 &  3.8 G & \\
111028 & K1III-IV & S    & 3.610 & 2.397  & 1.717  & 4.34  & 1.8 &     &   \\    
113226 & G8IIIab  &   S  &  3.694 & 0.37 &  0.089 & 5.48  & 2.3  &  &    \\   
115659 & G8IIIa    & S    &3.703 & -0.047 &  -0.29 &    5.70  & 3.6  &  4.2 GP & 
\\ 
116243 & G6II     &  S   & 3.697 & 0.037 & -0.22 &  5.12 &  4.2    &     & \\     
123139 & K0IIIb   &  S   & 3.663 & 0.70 &  0.27 &  5.11 &   1.2  &   & \\     
124897 & K1IIIbCN1 &   S & 3.622 & -0.31 & -0.99 & 4.53 & 1.5  & 2.4 G & \\ 
150798 & K2IIb-III & S   & 3.605 & -4.23 & -4.90 & 5.19 & 6.4  &   & 0.13 \\
151680 & K2.5III   & S   & 3.622 & 0.78  &  0.12  &4.83  &1.2   &2.2 GP & \\  
157999 & K2II    & SB   & 3.628  &-4.66 & -5.15  & 5.45	& 4.2 &       & \\
159181 & G2II    & SB   & 3.732  & -2.90 & -2.94 & 6.59 & 11.6 & 7.3 GT \\
162076     & G5IV    & S    & 3.653 & 1.27  & 0.82  & 5.45  &2.9  &     & \\
164058 & K5III   & S       & 3.586 & -1.04  & -2.03 & 4.50  & 1.2  &    & \\   
168723 & K2IIIabC1 & S     & 3.700 &1.84  &1.59   &5.23   & 1.2   &2.7 GN & \\
171443 &K3III-IIb & S      & 3.604 &0.21 &-0.61    &4.77   &1.8     &   & \\    
173764 & G5IISBO  & S   & 3.689 &-2.88  &-3.06  &6.15    & 8  &   & 0.1  \\
200905 &K4-5Ib-II & SB  & 3.580 &-4.07  &-4.83  &4.75  & 3.4  &  1.6 GT & \\ 
204867 & G0Ib     &  S  &  3.748 &-3.94 & -3.94  & 6.59 & 9.5  &  6.3 GT &  0.1 
\\
206778 & K2Ib     &  S  &  3.636 &-5.13 & -5.57 & 5.80 & 5.6  &  6.5 GT  & 0.2 
\\
209750 & G2Ib     &  S  & 3.732 &-4.26 &-4.30  &6.40  &7.8   & 6.7 GT   & 0.08\\
211416 & K3III    & SBO & 3.593 &-1.05  &-1.97  &4.63  &1.9    &     & \\     
218356 & K0II     &  SB &  3.668 &-2.45 & -2.72 & 6.20  & 1    &  3.9 GT &  0.24 
\\
            \hline
        \end{tabular}
	 \end{center}
   \end{table*}

\begin{table*}
\caption[]{Cluster data. NCC2516 photometry is from 
Lindoff (1973), NGC 6067 photometry from Thackeray et al. (1962)
IC4651  photometry from Anthony-Twarog et al. (1988), numbers from 
Lindoff (1972). For reddening and distance modulus, see text.
In column 11 V/R gives the ratio between the Violet and Red K2 emission in the
line core. V/R$<$1 may indicate outflows}
\label{table2}
\begin{center}
\begin{tabular}{lrrrrlllrrrrrrr}\hline
\multicolumn{1}{c}{ID}  &  
\multicolumn{1}{c}{B-V} &
\multicolumn{1}{c}{B-V$_o$} & 
\multicolumn{1}{c}{V-R} &
\multicolumn{1}{c}{V-R$_o$} & 
\multicolumn{1}{c}{T$_{eff}$} &
\multicolumn{1}{c}{M$_v$} &
\multicolumn{1}{c}{A(K)} & 
\multicolumn{1}{c}{A(50)} & 
\multicolumn{1}{c}{V/R} & 
\multicolumn{1}{c}{FKP} & 
\multicolumn{1}{c}{FKL} &
\multicolumn{1}{c}{F'$_k$} & 
\multicolumn{1}{c}{Vsini} & 
\multicolumn{1}{c}{Clus} \\
                       \hline
HR3120  &  1.55	& 1.43	& 1.03	& 0.95	& 3.620 & -2.61 & .250	& .0150	
& 1.02	& 5.00	& 5.22	&  5.07 & 2.8 & 2516 \\ 	
HR3153  &  1.74	& 1.62	& 1.44	& 1.36	& 3.554 & -3.17 & .884	& .0315	
& .21	& 4.41	& 4.41	&  4.35 & / & 2516  \\
\hline
CPD-537400 & 1.10 & .79 &      & (.55)	& 3.759 & -3.85 & .239 	& .0119	
& .83	& 6.22	& 6.35	&   6.19 & / & 6067 \\
CPD-537344 & 1.78 & 1.47 &     & (1.02)& 3.595 & -3.3  & .613
& .032 & .89 	& 5.17 & 5.33 	&  5.23 & /& 6067 \\  
CPD-537416 & 1.55 & 1.24  &    & (.84)	& 3.628 & -2.1  & .300	& .0157	
& .93	& 5.42	& 5.57	&   5.46 & / & 6067 \\
\hline
HD73598  &.96	& .96	& .72 & .72 &  3.672 & .03 & .120   & .0063  &  
/     & 5.39	& 5.55	&  5.37 & /& Praes. \\
HR3428  & 1.02	& 1.02	& .74 & .74  & 3.666 & .14 & .103  & .0048   & 
1.40  & 5.27	& 5.37	&  5.19 & 3.1 & Praes. \\
HR3427 &  .98	& .98	& .72 & .72  & 3.672 & .27 & .094 & .0050   & /  
  & 5.29	& 5.45	&   5.24 & 1.1 & Praes.  \\
\hline
HR1346	& .99   &  .99	& .73 & .73  & 3.669 & .28    &.159  & .0079   & 
1.14  & 5.49	 & 5.62 & 5.47  & 1.0 & $\gamma$Tau \\
HR1373	& .98	 & .98	& .73 & .73  & 3.669 & .41   &  .081  & .0044   
& 1.22  & 5.19	& 5.36 & 5.12  & 1.2  &  $\delta$Tau\\
HR1409	& 1.01  & 1.01 &.73 & .73  & 3.669 & .15   & .075  & .0042   & 
1.13  & 5.16 & 5.34& 5.08  & 2.4  & $\epsilon$Tau  \\
HR1411	& .95	& .95	& .71 & .71  & 3.675 & .42   & .166& .0082   & 
1.0   & 5.57	& 5.69	&  5.56  & 1.5  & $\theta$Tau  \\
\hline
IC4651-56 & 1.65  & 1.56   &     &(1.28)  & 3.554 & -.98 & .448 & .0210 & 
.91 & 4.23 & 4.35 &   4.23  & / \\
IC4651-37 & 1.03& .95	 &     & (.70)	& 3.674 &  .22 & .082 & .0040 & 
/   & 5.29 & 5.41 &   5.19   & /&  \\
IC4651-113 & 1.13& 1.05   & 	&(.77)	& 3.658 &  .37 & .092 & .0044 & 
.99 & 5.12	& 5.4	& 5.02   & /& \\
             \hline
         \end{tabular}
	\end{center}
    \end{table*}

  \begin{table*}
      \caption[]{New Ca II observations of field stars. Column 1-3 as in 
Table 1; Column 4: Measured (V-R); Column 5: Adopted (V-R)$_o$; Column 6: 
Effective Temperature; Column 7: M$_v$ (Hipparcos)
Column 8: M$_{Bol}$; Column 9: Ak Index; Column 10: A(50) 
Index; Column 11: K line V/R 
asymmetry; Column 12: Log Ca II K line chromospheric flux (ergs cm$^{-2}$ 
sec$^{-1}$ at the stellar surface); 
Column 13: {\it Vsini}(CORAVEL); Column 14: {\it Vsini}(Gray)}
         \label{table3}
         \begin{center}
	\begin{tabular}{l|l|l|r|r|r|r|l|l|l|r|r|r|r}\hline
\multicolumn{1}{c}{HD}  &  
\multicolumn{1}{c}{Sp. Ty.} & 
\multicolumn{1}{c}{Bin} & 
\multicolumn{1}{c}{(V-R)} &
\multicolumn{1}{c}{(V-R)$_o$} & 
\multicolumn{1}{c}{T$_{eff}$} & 
\multicolumn{1}{c}{M$_v$} & 
\multicolumn{1}{c}{M$_{Bol}$} & 
\multicolumn{1}{c}{A(K)} & 
\multicolumn{1}{c}{A(50)} & 
\multicolumn{1}{c}{V/R} & 
\multicolumn{1}{c}{LogF'$_k$} &  
\multicolumn{1}{c}{VC} & 
\multicolumn{1}{c}{VG} \\
 \hline
496   & K0III  &    S  & 0.75 &  0.75 &  3.665 & 0.71 &  0.28 &   .095 &  .0046  &  1.2 &   5.09 &  1.5 & 2.9 \\
30834 & K2.5III &   S & 1.09  & (0.96) & 3.618 & -2.00 & -2.68 &  0.190 & .0110    & 1.04 &  4.94  & 2.7  &    / \\
47205 &  K1III  &   S  & 0.79 & 0.79   & 3.655 & 2.46  & 1.99  & .080 &.0045 & 1.07 & 4.92   &  1 &   2.7 \\   
57623 & F6II    &   S &  0.67 & (0.45) & 3.787 & -3.60 & -3.52   & .345 & .018 &  1.23 &  6.73 & 11.8 &   /  \\
62509 & K0III   &   S  & 0.75 & 0.75   & 3.662 & 1.09  & 0.66   & .077 &.0043 &1.08  & 5.02  & 1.2  &  2.5 \\  
67228 & GIV     &   S  & 0.63 & 0.63   & 3.719 & 3.46  & 3.29  & .071 & .0031 & 1.11 &  5.3 &  2.9 &   3.7 \\ 
78647 & K4Ib-I  &   S  & 1.24 & 1.24   & 3.570 & -3.99 & -5.19 & .743       & .0350 &  0.57 & 4.6  & 8.9  &  /    \\ 
80230 & M1III   &   S  & 1.34 & 1.34   & 3.542 & -1.74 & -3.4 & .434 &.0205 &  0.44 & 4.14 & 3    &   /   \\
89484 & K1III   &   SB & 0.85 & 0.85   & 3.636 & -0.92  & -1.49 & .116 & .0067 & 1.0  &  4.95 &  1.1 &  2.6 \\
94481 & G4III   &   S  &  /  &(.69)   & 3.676  & 0.16 & -0.21 & .106 & .0052 & 1.0  &  5.37 &  2.8 &   /  \\
101379 & G2III+A & SB & 0.79 & 0.79   & 3.655  & -1.17 & -1.66 & .316 & .0084  & /   & 6.52  & 12.9 &   /  \\
101570 & G3Ib   &  SB & 0.87 & (.63)  & 3.714  & -2.99 & -3.09 & .962  & .0402 & 1.06 & 6.58  & 21.4  &  /  \\
118219 & G6III  &  S  &  /   & (.70)  & 3.672  & 0.42 & 0.05 & .084 & .0043 & 1.1  &  5.22  & 1.1   &  /  \\
121107 & G5III  &  S  &  /   & (.69)  & 3.676  & -0.89   & -1.27 & .193 & .0090 &  /   &  5.68  & 14.5  &   / \\
124850 & F6III  &  S  & 0.5  & 0.5   & 3.754   & 2.42  & 2.33  & .155 & .0057 & 1.06  & 6.11  & 15.6   & 15  \\
126868 & G2IV   &  SB & 0.58 & 0.58  & 3.732   & 1.73  & 1.59  & .320 & .0130 &  /    & 6.23  & 14.3   & 14.3 \\  
130259 & G5III  &  S  &  /  & (.69)& 3.676 & 0.20 & -0.18 & .253 & .0118 & 1.02  & 5.82  & 4.9    & /    \\
138716 & K1III-IV & S & 0.77 & 0.77 & 3.658 & 2.30  & 1.85  &.080 &.0043  &1.09  & 4.97  & 1.4   &  2.5 \\ 
140573 & K2III  & S  & 0.81 & 0.81  & 3.650  & 0.87 & 0.38 & .081 & .0049 & 1.13 & 4.87 & 1.9   &  0.  \\  
142980 & K1IV   & S  &  /  & (.82)  & 3.647  & 1.33 & 0.84 & .085 & .0049 & .99  & 4.87 & 1.5  &  1.1  \\ 
144608 & G3II-III &  S  &  0.65 &0.65 &3.711 &-0.24  &-0.44 & .091 & .0045 & 1.16  & 5.42 & 1.4   & /    \\
148856 & G7IIIa   & SBO&0.64 &0.64 &3.715  &-0.50  &-0.70 &.097 &.0049 &1.31   & 5.4   & 4.8   & 3.4  \\  
161096 & K2III    &  S & 0.82 & 0.82 &3.647  & 0.76  & 0.27 &.085 &.0052 & 1.15  & 4.9   & 1.2   & 1.6  \\ 
168723 & K2III    &  S   & 0.70 & 0.70 & 3.672 & 1.84 & 1.47 &.074 & .0038 &1.13  &5.14  &1.2    & 2.7  \\
181391 & G8III-IV &  SBO &/   & (.70) & 3.672 & 1.61  & 1.24 &.149 &.0069 &1.08  &5.63  &2.8    &2.7  \\ 
182572 & G8IV     &  S    & /  & (.66) & 3.705 & 4.27   & 4.05  &.080 & .0037 & /   & 5.29 & 1.6   & 2.3 \\
188512 & G8IV     &  S   & 0.66 & 0.66 & 3.705 & 3.03 & 2.81   & .075 &.0035 &1.1  & 5.26 & 1.2   & 2.2 \\
192876 & G3Ib     &  SB & 0.79 & (.63) & 3.714 &-3.07 &-3.16 & .493 & .0239 &1.07 & 6.32 & 7.3   & 6.2 \\  
196725 & K3Ib     &  S   &1.02  &1.02  & 3.598 & -2.32 & -3.07 &.448 &.0236 &.33  & 5.08 & 2.9    &0.0 \\ 
196755 &G5IV+K2V  & S  & /   & (.70) & 3.672 & 2.68  & 2.31  &.111 & .0050 & 1.1  & 5.35 & 3.5   & 3.3 \\ 
203387 & G8III  &  S    & 0.62 &0.62 & 3.716 & 0.18  & -0.02 & .322 & .0147 & 1.08 & 6.14 & 5.1   & 6.4 \\ 
206859 & G5Ib   &  S  & 0.80 & (.67)&3.698 &-3.48 & -3.62 &.411 &.0205 & 1.00  & 6.13  & 5.5    & 5.7 \\   
204075 & G4Ib   & SB & 0.64 & 0.64 &3.712 &-1.66  &-1.76  &.402 &.0189      & 1.22   & 6.19  & 3.8   &  6.2  \\  
67523  & F6II     & /    & 0.35    &  0.35 &      & 1.41   &   &      &       &      &       & 14.6 &  / \\
78791  & F9II     & /    & 0.55    &       &      & -1.25  &   &      &       &      &       & 66.5: &/ \\
79940  & F5III    &  /   & 0.38    & 0.38  &      & 1.10   &   &      &       &      &       &  /    & / \\ 
96918  & G40-Ia   & /   & 0.87     &       &      &  -7.34 &   &      &       &      &       &   /   &  /\\ 
100261 & G30-Ia   &  /   &  0.83   &       &      &  -4.91 &   &      &       &      &       &  /    & / \\ 
101947 & G00-Ia   &  /   & 0.76    &       &      & -4.48  &   &      &       &      &       &  /    & /  \\ 
129502 & F2 III   &  SB  & 0.40    &  0.40 &      & 2.51   &   & &       &    &       & 4.9/11.2     & /  \\ 
155203 & FIII-IVp &  /   &  0.36   & 0.36  &      & 1.61   &   &      &       &      &       &   /   & /  \\ 
174474 & A2V      &  /   &  /      &       &      &  1.42  &   &      &       &      &       &    /  & /  \\
178524 & F2II     &  /   & 0.34    & 0.34  &      &  -2.77 &   &      &       &      &       &  3.2 &  / \\
182835 & F2Ib     &  /   & 0.51    &       &      &  -8.1  &   &      &       &      &       & 11.5 &  / \\
196524 & F5IV     &  /   & 0.40    & 0.40  &      & 1.26   &   &      &       &      &       & 49.8 & /  \\
\hline
\end{tabular}
\end{center}
   \end{table*}

\end{document}